# Safety on Judo Children: Methodology and Results

***Attilio Sacripanti:*** *EJU Scientific Commission, FIJLKAM*
***Tania de Blasis****: University of Tor Vergata, FIJLKAM*

Biomechanics, Children Safety; Judo Throws, Elastocaloric Effect, Impact Biomechanics, Crash Test Methodology, Thermo mechanics of materials.

The main problem against the acceptance of judo for children as sport, in the families, is the "strange" position that some medicine doctors have respect to judo. Many doctors although they have not firsthand experience of judo, describe it as a sport unsuitable for children, often expressing themselves so broadly negative, even via web. Theoretically speaking falls derived by Judo throwing techniques, could be potentially dangerous, especially for kids, if poorly managed. Obviously all judo people knows that good judo is safe for children, and *how these affirmations are generalist and negative*, but the "truth" is based only on personal experiences, not supported by scientific evidence worldwide accepted.

A lot of researches were focalized on traumas or injuries taking place in judo, both during training and competition, from these papers, you know, paradoxically, that training has a higher content of incidents against competition. However never a totally complete Scientific Research was performed to support the harmlessness of judo for kids, especially with regard to potential traumas deriving by falls due to throwing techniques. The goal of this Research is to define and apply a scientific methodology to evaluate the hazard in falls by judo throws for children during training. By organizing the research on the basis of Safety Analysis there are at first defined and experimentally evaluated for fifteen among boys and girls:

1. The flight time for five throws
2. The maximum impact forces and velocities for five throws
3. The contact surfaces of bodies on the Tatami for five different throws.
4. The Elastocaloric effect to evaluate the energy absorption by Tatami Materials.
5. The maximum Strain on the Tatami.
6. The impact reaction Stress on the children bodies.

After valuated the mechanic of falls, and the evaluation of contact body surfaces on the Tatami, by a Japanese AVIO Thermal Camera, the next step is to connect the impact biomechanics with the potential traumas. The only worldwide accepted methodology both from Medical and Engineering people is the Crash test Methodology. Along with the appropriate changes in the specifications of the "Crash Test Methodology" there are defined and evaluated:

1. A "judo boy Dummy", to apply safety criteria used in crash test.
2. Both: Thoracic Trauma Index and Compression Index
3. The Head Injury Criterion
4. The probability of skull fracture (if any) applying risk analysis.

Connecting in this way, the mechanical results with the resulting physiological hazard connected to Judo throwing techniques, using the "Crash Test Methodology" it is proved that, correct falls of judo throws are safe for "judo boy Dummy", and for logical extension they are safe, also for judo children.

The last two troubles of still judo throwing techniques training for kids are:

The wrong combination of kids during training and the potential long term traumas, in such delicate issues a Digital Assistant named (Hazard Training Sentinel) have been prepared to help teachers to manage in optimal way this specific aspect of their important and delicate work.

# Safety on Judo Children: Methodology and Results

*Attilio Sacripanti:* EJU Scientific Commission, FIJLKAM
*Tania de Blasis*: University of Tor Vergata, FIJLKAM

# Safety on Judo Children: Methodology and Results

*Attilio Sacripanti:* EJU Scientific Commission, FIJLKAM
*Tania de Blasis:* University of Tor Vergata, FIJLKAM

## 1) Introduction

The perennial issue is :
Are Judo Throws really dangerous for trauma in children ?
Into the web or into some books it is possible to find information like that:
*"Because Judo involves throws, flips and takedowns, it has a higher rate of injuries to the upper extremity. Injuries to the shoulder, elbow, wrist, hand and fingers are common."…"Kids in judo suffer more shoulder, upper arm and neck injuries than kids in karate or taekwondo"*[1]
Theoretically speaking falls produced by Judo throwing techniques, could be potentially dangerous, especially for kids, if poorly managed.

**Fig 1** *Injuries in function of year and sex (children) [2]*    **Fig2** *Injuries in training and competition [3]*

Obviously all judo people knows this situation [4] and teachers are very attentive to teach the ukemi, whether alone or connected to judo throws, especially to children.
But, it is also very clear that *affirmations like the previous ones' are generalist and negative*, and they are the main obstacle to increase the number of enrollment of new children in the Dojo (gyms) of the world.
Against these doctoral affirmation, the judo teacher knowledge is based only on various personal experiences, and never a totally complete Scientific Research supported this knowledge.
Why? Because it is very difficult to evaluate all main parameters that allow these experiences in scientific way. Since they belong to many interconnected scientific fields, among all:
Physics, Mechanics of Composite Materials, Thermodynamics, Biomechanics of trauma, Forensic Medicine, Biology, Bioengineering of Crash Test , Risk Analysis, Safety Science, and last but not least, the most complex one : JUDO.

## 2) System Safety Criteria

The basic philosophical approach, in the "situation" which must be studied: *safety for falls due to throws of children's judo*, leads us to evolve the concept of safety, normally accepted into safety engineering world.

In this field the safety concept requires a risk management strategy based on identification of hazards and application of remedial controls using a systems-based approach. [5]

In our situation "Falls produced by Judo throwing techniques" the hazard is the condition that can cause injury to children thrown by judo techniques.

Obviously our study is not focalized on the ukemi poorly managed, but on the normal situation: children and kids that throw each other, with their "beginners" knowledge about ukemi and throws.

Normally, people think of safety as the absence of accidents and incidents (or as an acceptable level of risk). In this perspective, Hollnagel, Wears and Braithwaite define this common vision: Safety-I.

In our situation, safety-I could be defined as a state where *as few ukemi as possible go wrong*.

A Safety-I approach presumes that ukemi go wrong because of identifiable failures of specific situations. It was tacitly assumed (in engineering practice) that the "situation" analyzed could be decomposed into steps and such steps could be improved.

As "situations" continue to develop more complex, adjustments become increasingly important, to maintain a safe outcome, but they become also more complex to understand.

The challenge for safety improvement is therefore to understand these adjustments, in other words, to understand how outcome usually goes right.

Despite the obvious importance of ukemi going right, traditional safety management has paid little attention to this.

Our Safety management should therefore move from ensuring that '*as few ukemi as possible go wrong*' to ensuring that '*as many ukemi as possible go right*'. Hollnagel, Wears and Braithwaite call this perspective Safety-II. [6]

All this philosophical talk flows for us, as practical aspect, into the analysis of 90th percentile of Ukemi's Gaussian, studied by the right side limit, that is the area of ukemi "*that go right*", to understand if they are risky for children in every day conditions.

## 3) *Mechanics of Throwing Techniques and Impact Biomechanics*

It is meaningless for this specific work to analyze the mechanics of throwing techniques, because, for each one, there are infinite time-varying parameters. [7][8][9][10][11][12][13][14][15]

Even trajectories are infinitely variable, for example in one Japanese study on breackfalls, we can find expression like the following: "The kinematic data of the breakfall motion for both Osoto-gari and Ouchi-gari were collected using a three-dimensional motion analysis technique (200 Hz). It is observed significant differences between the movement patterns for the two techniques, especially in the lower extremity movements." There is the need to change approach to our problem if we like to obtain significant, valuable and general solutions. Then to study each technique in his specific mechanical aspects is meaningless because variation are infinitive, and a statistical approach also not good. However the instant of impact and its biomechanical results is the only reproducible time with traumatic meaning. Its mechanics is concerned with reaction forces that develop during the collision and with the dynamic response of structures to these reaction forces.

This subject has a wide range of engineering applications, from designing sports equipment to improving the crash worthiness of automobiles. His analytical methods of solution in our situation derive by the simple Newton collision theory of rigid bodies.

Generally speaking, medical people know that, falls are dangerous for human bodies, because more often the collisions are plastic collision of a "visco-elastic" body against a rigid one's, today there are mathematical tools that combine mechanics of contact between elastic-plastic bodies with dynamics of structural response. [16]

But how children are, potentially, more exposed in judo to danger of fall than adults ?

Because children's bodies are under development and therefore their bones, within certain limits, are more flexible this means a decreased capability to absorb energy and an increased danger for internal organs, the developing joints are not completed tendon insertion more fragile and their bodies in standing position are more unbalanced than men.

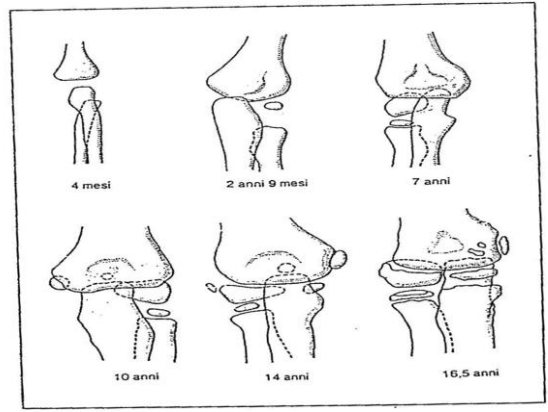

Fig 3 evolution of the elbow joint in time

In our case: child's body falling, thrown by a judo technique, the biomechanical result connected to the impact instant is the only reproducible point with general traumatic meaning. [17][18]
Then at light of our safety approach criteria and remembering both the Principle of independence of simultaneous actions and the vectors property of to add together, it is significant to study the instant of impact in its worst conditions , as show in the following way.

## 4) Safety Criteria Mechanical Analysis of falls produced by Judo throws

If we are interested, at the impact instant, the general motion equation from Galileo is:

$$h(t) = \frac{1}{2}gt^2 + v(t) \qquad (1)$$

v(t) is the added velocity by Tori, to consider the worst conditions we can hypothesize that the added velocity at the contact instant on the Tatami is proportional, a part or a multiple of the final velocity of free fall (for h=1m ) equal to:

$$v_f(t) = k\sqrt{2gh} \qquad (2)$$

If we remember that collision is an impulsive phenomenon, the conservation of momentum and the Newton definition of force give us the capability to state the following equation:

$$\int_0^t F dt = \int_v^v m dv \Rightarrow \cong F = \frac{mk\sqrt{2g}}{\Delta t} + \frac{m\sqrt{2g}}{\Delta t} \quad \text{with } h \equiv 1m \qquad (3)$$

From the well know equation of the kinetic energy it is possible to evaluate the Maximum, time independent, Impact Force on the Tatami, when added velocity by Tori is, as hypothesized proportional to maximum impact velocity.
Then:

$$E_f = \frac{1}{2}m\left[(1+K)\sqrt{2g}\right]^2 \Rightarrow F_f = \frac{1}{2}\frac{m}{h}\left[(1+K)\sqrt{2gh}\right]^2 \quad for \ h = 1m \qquad (4)$$
$$F_f = (1+K)^2 mg \qquad (5)$$
And K=0,1,2,3…,n .

Now, remembering the Principle of independence of simultaneous actions and the vectors property of to add together, we can show both: how flight time and velocity and how velocity and Impact Force, are connected.

All this could be shown by a parametric curves depending on the k parameter. Then the Added up Velocity increases, Maximum Impact Force increases, and Flight Time consequently decreases. The two curves are valid for all judo players for whatever age and mass.

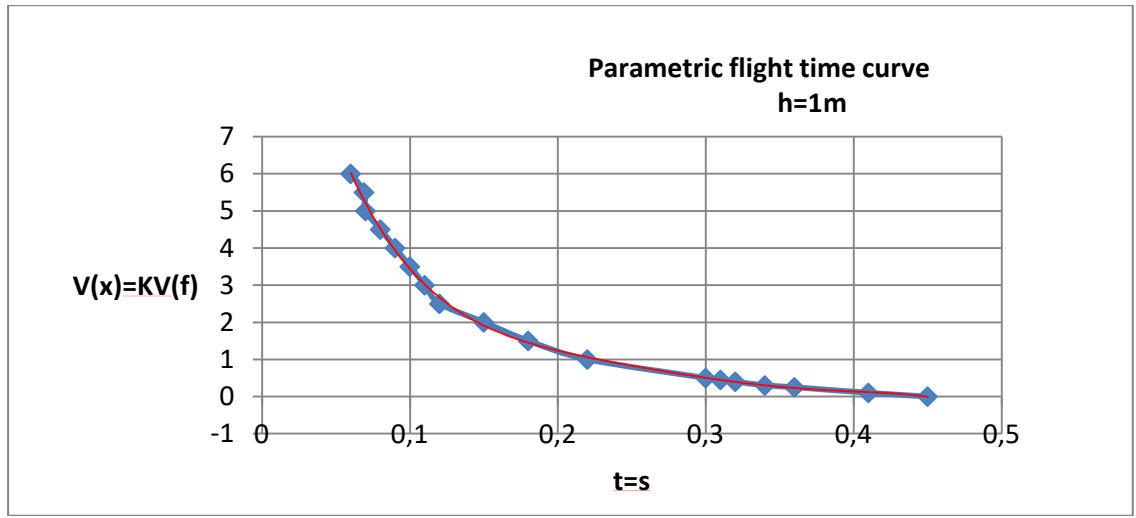

*Fig 4 Flight time vs velocity at impact*

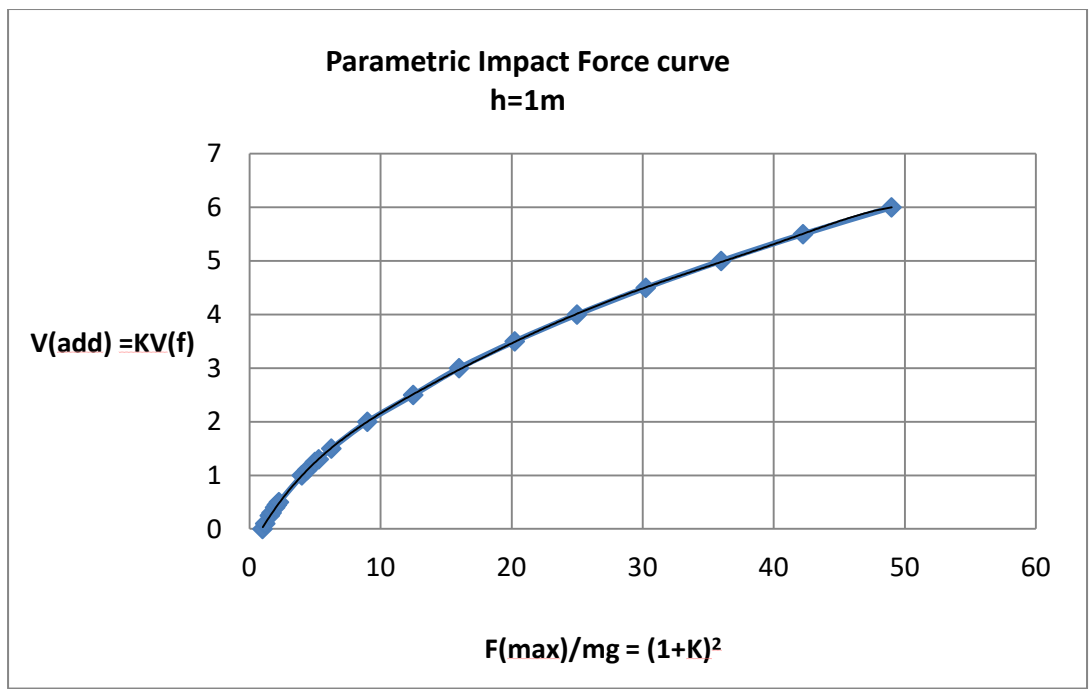

*Fig5 Maximum velocity at impact vs Par Maximum Force at impact*

From these two curves valid for every judokas, from children to high level competition, falling from 1 meter, it is clear that falls can be theoretically dangerous if the added velocity is high.

However we must remember that velocity added is directly proportional to force applied to throw the adversary.

For kids or children that throw each other, forces are small respect to high level competitors, and consequently should be small both the added velocity and the maximum impact force, as clearly show the next two diagrams.

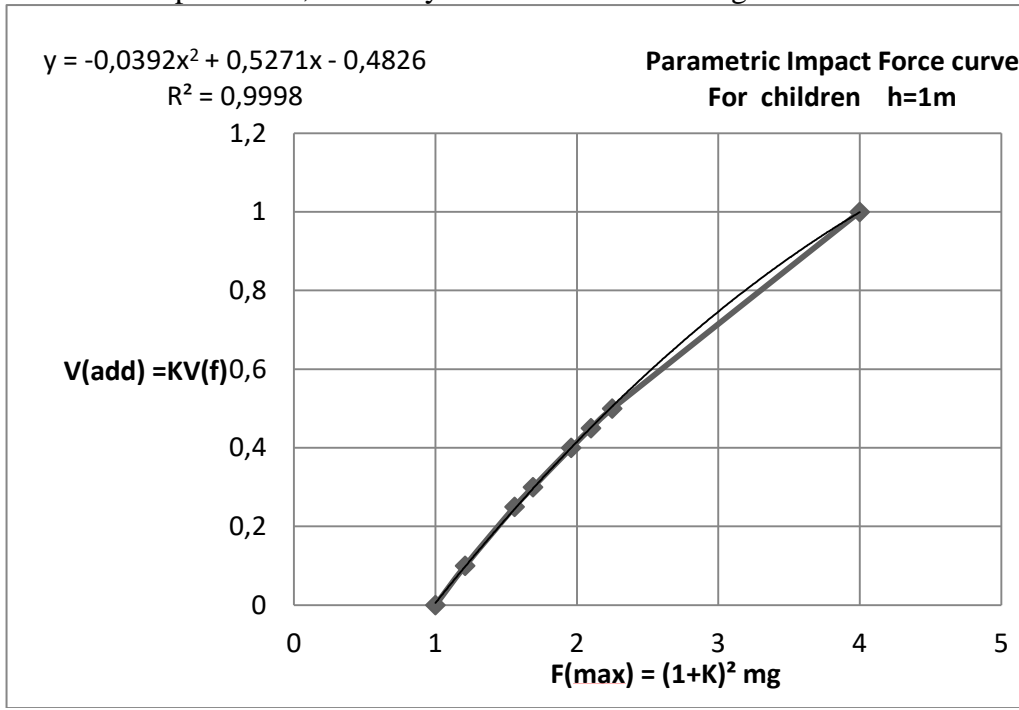

*Fig 6 Velocity vs Maximum Force at impact for children*

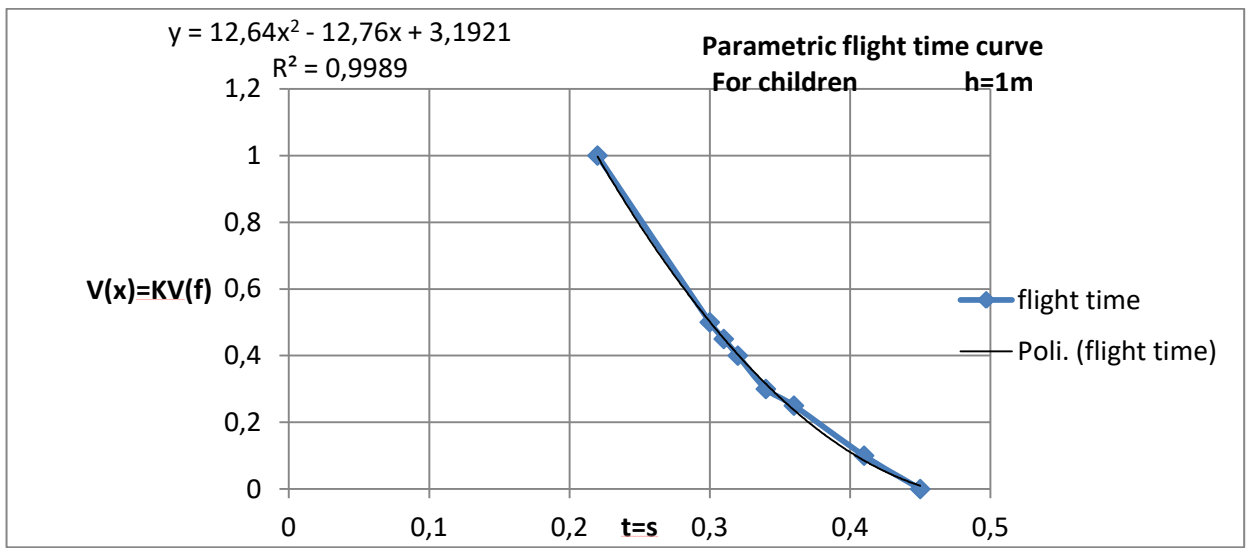

*Fig 7 Flight time vs velocity at impact for childre*

## 5) *Tatami Material Science and Thermodynamics.*

Material science is essential part of this research, because safety outcome depends both from the Tatami material and quality. In this research we analyzed one Tatami built by polyurethane foam and soft polyurethane covered by pvc and Approved by IJF, with thickness of 4 cm, and overall density 240 kg/m³. tensile strength 2480 N/5 cm, theoretical Force reduction ≈ 25%-40%

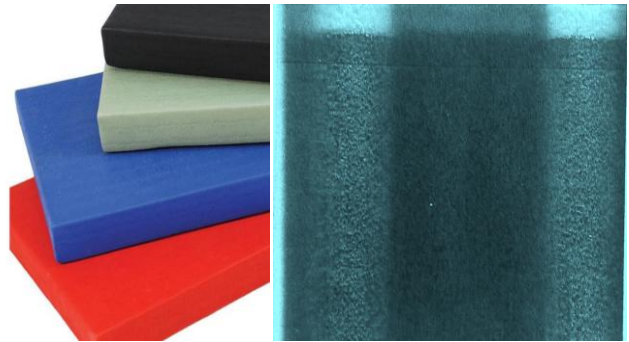

*Fig 8 Tatami and vertical constituents section*

PU invented by Bayer in Germany around 1937, have a history of slightly more than 75 years. They have become one of the most dynamic groups of polymers. Their use covers practically all fields of polymer application: foams, elastomers, thermoplastics, thermorigids, adhesives, coatings, sealants, and fibres.
PU are obtained by the reaction of an oligomeric polyol [low-molecular weight (MW) polymer with terminal hydroxyl groups] and a polyisocyanate.
The structure of the oligomeric polyol used for PU manufacture has a very profound effect on the properties of the resulting polymer, as assure us Ionescu in his encyclopedic work [19]. The tatami analyzed was built by three layers first layer PVC, second Polyurethane foam, and third Poliurethane semi rigid.
The foam is important but his mechanical evolution is quite complex.
The response of foam gets stiffer with increase in strain rate, and densification (lockup) occurs well below the strains at which lockup occurs for foam deformed at quasi-static strain rates. Consequently, the energy absorption characteristics of foam are altered with change in strain rate.[20]

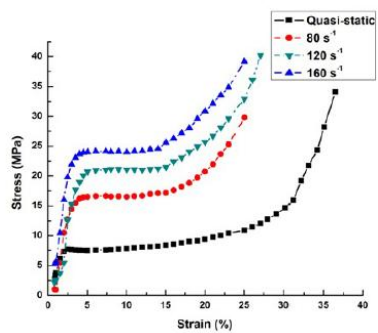

*Fig 9 Stress strain foam curves*

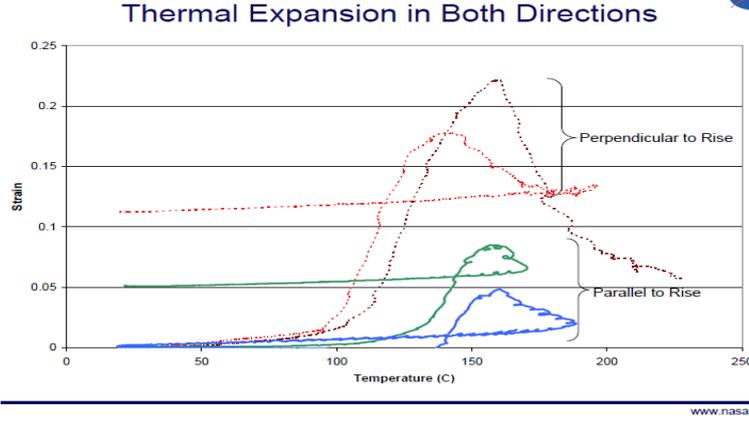

*Fig 10 Polyurethane Foam thermal expansion*

Also very complex is the thermal behavior of polyurethane foam, as NASA researchers shown in some very interesting works [21].

Since the foam is not a material, but a structure, the modeling of the expansion is complex. It is also complicated by the anisoptropy of the material. During the spraying and foaming process, the cells become elongated in the rise direction and this imparts different properties in the rise direction than in the transverse directions.

However we are much more interested than expansion, in his compression and related thermo-dynamical effects.

If the compression produced by children's body is fast, the situation can be approximated in thermo-dynamical terms to an adiabatic transformation.

This specific transformation was named by Viecheslav Sychev in his book "Complex thermodynamic System" [22] Elastocaloric Effect.

## 6) *Elastocaloric Effect*

When a body falls, on the tatami, after a Judo throw; the impact produces one adiabatic compression of the tatami, the impact energy will partially absorbed and one of the main effect, that changes the mechanical energy into heat, is the *Elastocaloric Effect*.

The induced variation of temperature is expressed by the following easy calculation:

$$T = T_0 + \int_0^{\Psi} \left(\frac{\partial T}{\partial \Psi}\right)_{S,P} \partial \Psi \quad (6)$$

To solve the kernel of the integral we can use the Maxwell equation

$$\left(\frac{\partial T}{\partial \Psi}\right)_{S,P} = -\left(\frac{\partial l}{\partial S}\right)_{\Psi,P} = -\left(\frac{\partial l}{\partial T}\right)_{\Psi,P}\left(\frac{\partial T}{\partial S}\right)_{\Psi,P} \quad (7)$$

And after few simple calculation we have the following final Relationship:

$$T = T_0 - \frac{\alpha_l \overline{T}}{c_p \rho}\Psi \Rightarrow \Delta T = -\frac{\alpha_l \overline{T}}{c_p \rho}\Psi \quad (8)$$

When the Tatami is compressed the stress $\Psi$ is negative and the Tatami temperature increases, absorbing energy.

The previous final relationship (8), remembering the Hookean Elastic Equation can be changed as:

$$T = T_0 - \frac{\alpha_l \overline{T}}{c_p \rho} \Psi \Rightarrow \Delta T = -\frac{\alpha_l \overline{T}}{c_p \rho} \Psi = \frac{\alpha_l \overline{T} E \Delta \varepsilon}{c_p \rho} \quad (9)$$

To have a first indicative order of magnitude in our research, very simple "theoretical" evaluation, assures, that with Polyurethane Foam as Tatami material IJF Licensed, with **density 244 Kg/m³**, with energy absorption, **around 25%- 35%**, hypnotized *Theor. Compr* ≈ 2 mm, temperature will have a "theoretical" increase of : **296.15<T(°K) <297.0** or in Celsius **23 < T(°C) <23.8**
Mechanics and Elastocaloric effect are connected by means of Strain that Produces Tatami Compression by Hook law. Compression is produced by Children body falling down and part of this Strain, after energy absorption, is returned to the children body for the Action Reaction Principle. In formulas:

$$\Psi = \frac{F}{A} = E \Delta \varepsilon \Rightarrow \Delta \varepsilon = \frac{\Delta l}{l} \quad (10)$$

$$\Psi' = -e\Psi \quad (11)$$

In which ( **e** < 1) is similar to the restitution coefficient and depends from the Tatami Material.
Remembering that : $\Psi' = -e\frac{F}{A}$ \quad (12)

**A ( The Children Body Impact Surface)** is the key parameters that we ask to find in this research to evaluate potential trauma in children produced by falls of judo throws.
How to evaluate this key parameters, by means of thermal image of the children body surface of contact with the tatami after the falls produced by a judo throw.
This phenomenon is well known in thermodynamics, and it is similar to an iron burning, it is a problem of transient conduction from the children body to the contact surface that heat the Tatami surface layer, and if the children go away fast it is possible to fix with a good sensible thermocamera the image of contact body area during a convective cooling.
A possible example equation of this phenomenon is presented with the some differences in the well known book, of Professor Latif [23]. Modeling the skin-layer of Tatami as an insulated plate of thickness L, conductivity k and convention coefficient α, using Duhamel's integral to determine the one dimensional transient temperature of the skin-layer, a solution for constant heat flux will be:

$$\overline{T}(x,t) = \frac{L}{k}\left[\frac{3x^2 - L^2}{6L^2} - \frac{2}{\pi^2}\sum_{n=1}^{\infty}\frac{(-1)^n}{n^2}\cos(n\pi x / L)e^{-\alpha(n\pi/L)^2 t}\right] \quad (13)$$

However we don't need a Thermo-dynamical numerical solution for our research but an arithmetical one connected with the numerical evaluation of contact surface area, for this purpose the Tatami surface was divided in squares of 10 cm x 10 cm. as in the following figures that show the set up preparation.

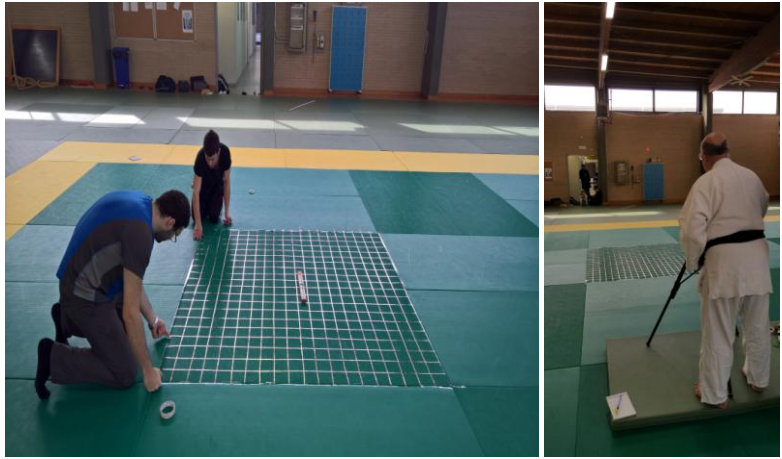

*Fig 11 Experimental set up preparation*

.

## 7) *Judo Throws and Their Specific Way of Falls, in Safety Analysis*

In our research considering as main parameter, from a safety point of view, the direction of free fall velocity, standing judo throwing techniques, also considering the beginner training situation, have several ways to fall, depending from the general mechanics of the technique.

The safety approach results in having to assess differently, from the normal approved convention, the flight times of projection trajectories. As exemplification of our safety approach, we consider two major throwing techniques:

Standing Ippon Seoi Nage and Tai Otoshi these two throwing techniques in our point of view show two different flight times assessments.

For standing Ippon Seoi, because added velocity takes equal direction to gravitational acceleration only towards the second half of the trajectory,

While in Tai Otoshi the vector direction is concordant since the beginning of the trajectory Considering in our approximation a complete time flight: the time that Uke detaches his feet from the tatami till to his body lands on the tatami.

We must consider for Seoi Nage only the useful time to add the velocity to the fall, this means to consider only half time of flight in our approximation, while for Tai Otoshi the complete flight time is useful to add velocity to the final landing.

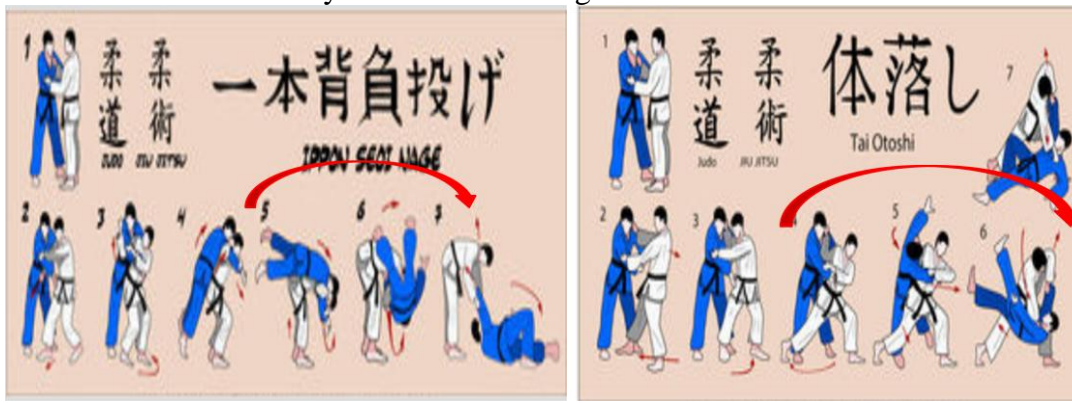

*Figg 12 13 Examples of times useful for the safety evaluation of the falls of two throws.*

On the basis of previous evaluations all most important throwing techniques for children were divided in three main groups: Lever, Couple, and Makikomi (in which Tori uses his weight to throw ) in such way all potential danger for Ukè will be covered by our safety analysis. Remembering that Sutemi for kids and children are not widely used.
Then we had analyzed :

| | | |
|---|---|---|
| For Lever Group | Ippon Seoi Nage | Half flight time |
| | Tai Otoshi | Complete flight time |
| For Couple Group | Uchi Mata | Half flight time |
| | O Soto Gari | Complete flight time |
| For Makikomi | O Soto Makikomi | Complete flight time |

## 8) *The Rational Goal of the Research*

The soul of this research on safety for judo children about falls produced by judo throwing techniques, is to define and apply a scientific methodology to evaluate the hazard in falls by judo throws for children.
Then as summary we evaluate and define:
1. The contact surface of children bodies with Tatami for five different throws.
2. The Elastocaloric Effect to evaluate the energy absorption by Tatami Material.
3. The impact reaction force on the children body
4. A "judo boy Dummy", to apply safety criteria used in crash test.
5. The probability of skull fracture ( if any) applying risk analysis .
6. Both: Thoracic Trauma Index and Compression Index.
7. Finally if correct falls of judo throws are safe or not, for "judo boy Dummy", they are so, also for children.

## 9) *Avio Thermal Camera and impact surface measurement.*

The impact surface measurement are based on the dynamic heat conduction between children body and surface layer of the tatami.
However also radiation can contribute, for example easy theoretical calculations show that in our situation a children body radiates on one square meter of Tatami at 23 C° on the basis of Stefan –Boltzmann law more or less a radiant emittance of 430 W/m$^2$
in formulas

$$E_m = \varepsilon \sigma T^4 = (0.98)\ 5.67\ 10^{-8}\ (296{,}15)^4 \approx 427\ \text{-}\ W/m^2 \qquad (14)$$

Few studies using infrared thermography have been devoted to sports performance diagnostic and to sports pathology diagnostic.
It is well known that sports activity induces a complex thermoregulation process where part of heat is given off by the skin of athletes. As not all the heat produced can be entirely given off, there follows a muscular heating resulting in an increase in the skin superficial temperature. In particular, the IRT method will enable, in the long term, to quantify the heat loss.
Just to make an historical correction to the history of thermography in sports [24] in Italy during the period 1989 - 1993 many researches, with thermography, were produced by the author coordinating a joint research a group of researchers of three Italian Institutions ENEA, CONI and FILPJ ,to assess, for the first time, the oxygen consumption of judo throws by thermocameras.
A that time Thermal cameras were freeze by liquid Nitrogen.
These old researches were also remembered in the recent research publication [25].

The old researches were focalized to evaluate, not only, the difference of Energy consumption among two groups of throws Lever and Couple, but also to build one heat exchange equation men-environment to evaluate the mean overall oxygen consumption in real competition.

This actual research is focalized on the capture of the thermal image of surface body contact left by Uke after the fall of throwing and on its measure.

When children body falling touches the tatami it leaves one thermal track produced by dynamic thermal conduction.

This not visible, thermal track disappear very fast due to the cooling by convention of the surface layer, when the body leaves the tatami.

The research idea is to capture by a fast and sensible thermal camera this evanishing image of the contact surface, measure it and evaluate for safety point of view, the stress received by children body that is the maximum impact force divided the measure of evanishing thermal image of contact surface.

In the next four figures there are shown some thermal captures obtained by the Japanese Thermal Camera AVIO 600 from Nippon Avionics.

Equipped with the software InfReC Analyzer 9500

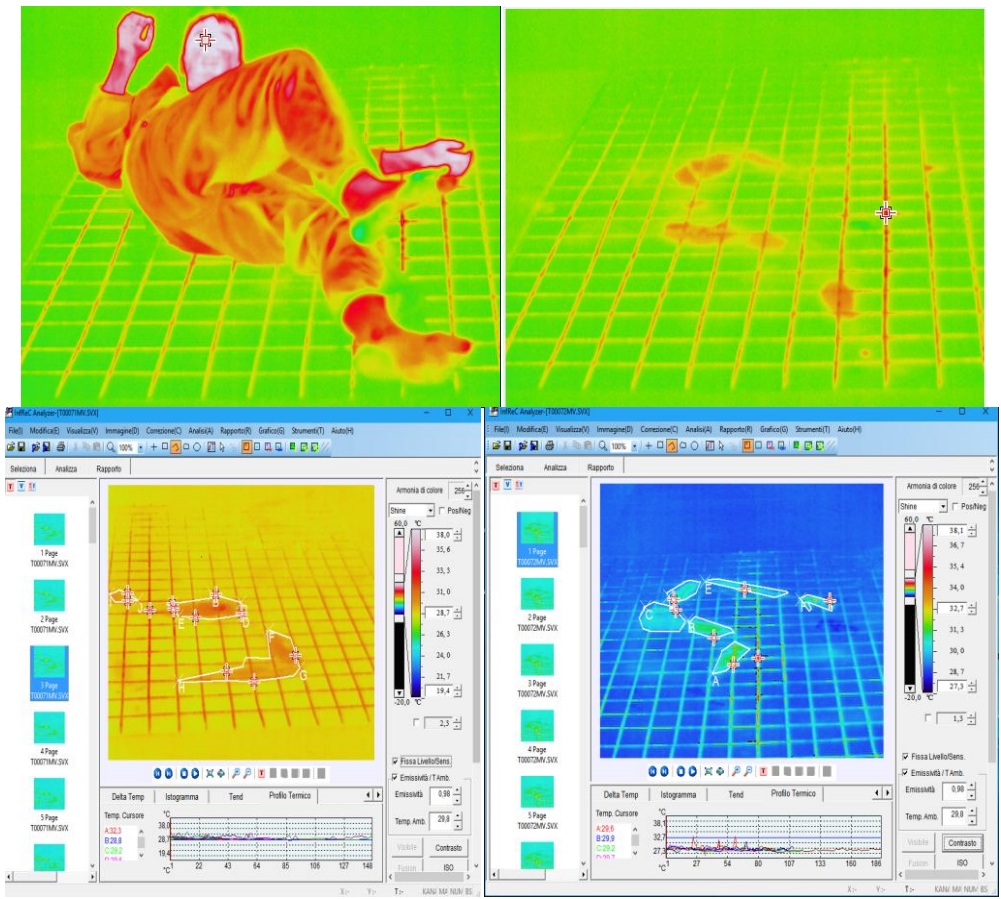

*Figg, 14-17 thermal capture of fall and track , and two thermal tracks bounded and measured.*

## 10) Data evaluating the Children Judo Maximum hazard for falls produced by judo throws: Mechanical and Thermal Information

In this paragraph are collected all the most important results of the research, these results are divided in three main areas

Source:

In the source we consider all the starting data of children analyzed, the thermal environmental situation of the FIJLKAM Dojo measured by an Rocktrail atmospheric station-

| Year | Height m | Weight Kg | thorax cm | SBA (Costeff) m² |
|---|---|---|---|---|
| 2003 | 1,47 | 39,7 | 75 | 1,278 |
| 2003 | 1,49 | 35,4 | 67 | 1,185 |
| 2002 | 1,64 | 58,6 | 63,1 | 1,624 |
| 2005 | 1,52 | 50,7 | 83 | 1,491 |
| 2002 | 1,61 | 52,5 | 73 | 1,521 |
| 2006 | 1,42 | 41,8 | 80 | 1,322 |
| 1998 | 1,56 | 54,3 | 82 | 1,553 |
| 2008 | 1,31 | 27,2 | 63 | 0,988 |
| 2001 | 1,77 | 65,8 | 90 | 1,734 |
| 2002 | 1,76 | 65,6 | 90 | 1,731 |
| 2001 | 1,84 | 98 | 100 | 2,122 |
| 2000 | 1,71 | 52,8 | 89 | 1,528 |
| 2002 | 1,46 | 49,4 | 80 | 1,468 |
| 2005 | 1.58 | 45,2 | 65 | 1,389 |
| 2004 | 1,69 | 66,1 | 84 | 1,739 |
| 2007 | 1,51 | 38,1 | 70 | 1,244 |

*Tab 1  Children data*

Barometric values: Environmental Temperature 25-27  C°

Internal Dojo Temperature 23-25 C°, Humidity  55%-57%

Alt. 0  Level of sea

Mechanical results

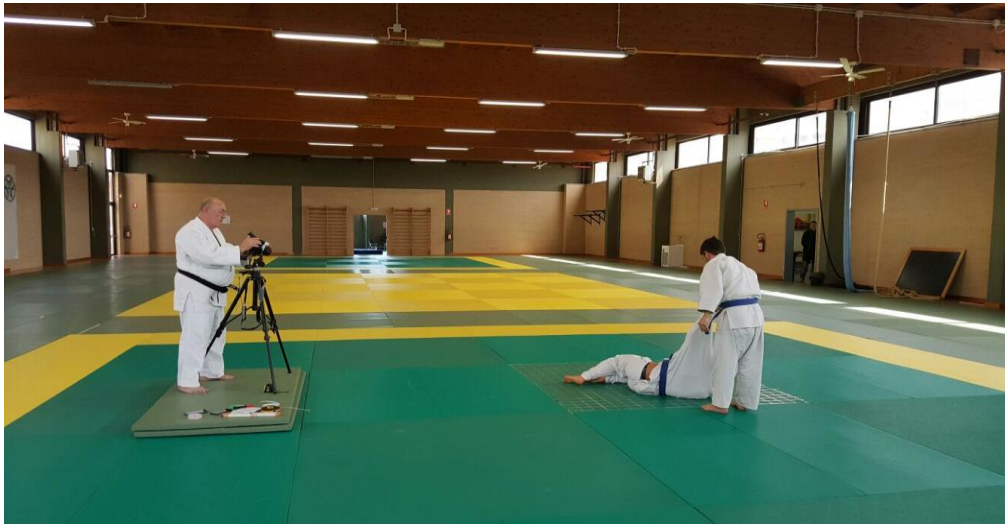

*Fig, 18  the experiment in visible light*

| Ippon Seoi Nage s | Tai Otoshi s | Uchi Mata s | O Soto Gari s | Soto Makikomi s |
|---:|---:|---:|---:|---:|
| 0,365 | 0,55 | 0,245 | 0,25 | 0,68 |
| 0,415 | 0,69 | 0,34 | 0,34 | 0,46 |
| 0,44 | 0,65 | 0,35 | 0,57 | 0,73 |
| 0,315 | 0,56 | 0,31 | 0,4 | 0,44 |
| 0,37 | 0,5 | 0,47 | 0,44 | 0,84 |
| 0,265 | 1,00 | 0,395 | 0,29 | 0,81 |
| 0,35 | 0,69 | 0,37 | 0,26 | 0,87 |
| 0,33 | 0,58 | 0,3 | 0,33 | 0,68 |
| 0,57 | 0,41 | 0,315 | 0,53 | 0,6 |
| 0,495 | 0,68 | 0,4 | 0,4 | 0,71 |

*Tab 2  Experimental flight time measured for 5 throws with a digital cronograph*

| Ippon Seoi Nage m/s | Tai Otoshi m/s | Uchi Mata m/s | O Soto Gari m/s | Soto Makikomi m/s |
|---|---|---|---|---|
| 0,22 | 0 | 0,82 | 0,79 | ≤0 |
| 0,07 | ≤0 | 0,31 | 0,31 | 0 |
| 0,02 | ≤0 | 0,27 | 0 | ≤0 |
| 0,43 | 0 | 0,45 | 0,11 | 0,02 |
| 0,2 | 0 | 0 | 0, 02 | ≤0 |
| 0,7 | ≤0 | 0,12 | 0,55 | ≤0 |
| 0,27 | ≤0 | 0,2 | 0,73 | ≤0 |
| 0,36 | 0 | 0,5 | 0,36 | ≤0 |
| 0 | 0,09 | 0,43 | 0 | ≤0 |
| 0 | ≤0 | 0,11 | 0,11 | ≤0 |

*Tab 3  Maximum Velocity parameter evaluated by fig. 7*

| Ippon Seoi Nage N | Tai Otoshi N | Uchi Mata N | O Soto Gari N | Soto Makikomi N |
|---|---|---|---|---|
| 1,5 | 1 | 3,28 | 3,16 | ≤1 |
| 1,15 | ≤1 | 1,74 | 1,74 | 1 |
| 1,04 | ≤1 | 1,64 | 1 | ≤1 |
| 2,03 | 1 | 2,1 | 1,24 | 1,04 |
| 1,45 | 1 | 1 | 1,04 | ≤1 |
| 2,84 | ≤1 | 1,27 | 2,39 | ≤1 |
| 1,64 | ≤1 | 1,45 | 2,94 | ≤1 |
| 1,85 | 1 | 2,24 | 1,85 | ≤1 |
| 1 | 1,18 | 2,03 | 1 | ≤1 |
| 1 | ≤1 | 1,24 | 1,24 | ≤1 |

*Tab 4  Maximum Force parameter evaluated by fig. 6*

*Mechanical results about hazard for children*

Max Impact Force: 1,55 mg, mg, 1,78 mg, 1,76 mg, mg
Max Impact Velocity (m/s): 5,39; 4,79; 5,81; 5,71; 4,48
Maximum Stress ( MPa) 0,13; 0,05; 0,09; 0,07; 0,05.
Max Stress Received by children body [25%](MPa) 0,09; 0,03; 0,06; 0,05; 0,03.
1 Atmosphere = 0,1 MPa
Rib experimental  max compression   0,0075 mm

## *Thermal results about hazard for children*

| Ippon Seoi Nage cm² | Tai Otoshi cm² | Uchi Mata cm² | O Soto Gari cm² | Soto Makikomi cm² |
|---:|---:|---:|---:|---:|
| 45 | 60 | 80 | 85 | 120 |
| 70 | 75 | 110 | 140 | 85 |
| 55 | 70 | 115 | 120 | 100 |
| 75 | 85 | 95 | 95 | 95 |
| 65 | 85 | 110 | 140 | 95 |
| 70 | 105 | 75 | 180 | 130 |
| 55 | 140 | 90 | 80 | 100 |
| 50 | 105 | 80 | 75 | 80 |
| 60 | 110 | 90 | 150 | 70 |
| 55 | 95 | 75 | 90 | 115 |

Tab 5 Contact Surfaces evaluated by thermal images ±7%

Maximum Stress ( MPa) 0,13; 0,05; 0,09; 0,07; 0,05.
Experimental evaluated Elastocaloric effect for five throws:

```
 In Kelvin                    In Celsius
296,15 <T(°K) < 296,17        23<T(°C) <23.02
296,15 <T(°K) < 296,16        23< T(°C) <23,01
296,15 <T(°K) < 296,168       23 <T(°C) <23,018
296,15 <T (°K) < 296,165      23 <T(°C) < 23,014
296,15 < T(°K) < 296.16       23 <T (°C) <23,01
```

These experimental measurements gives us the capability to evaluate the connected tatami compression which is around 0,5 mm !
2mm was the previous theoretical evaluation to have a reference measurement for temperature rise. This evaluation is one indirect measurement of the little intensity of forces applied among children that train judo throwing techniques.

## 11) *Data for "Judo boy Dummy" and Crash test methodology.*

One big problem in translate all the biomechanical and thermal calculation in physiological effects. In fact there is not a worldwide accepted way to connect impact Biomechanics data to the hazard for traumas in children.
After a large research [26] [27][28] [29] [30][31][32][33] the only way accepted in medical and engineering areas, in our opinion, is the **Crash Test approach**, considered sufficiently objective.
Following this approach, to objectify and generalize the preious judo results, we define a "**Judo Boy Dummy**", built by the mean data of the real boys, on which are applied the mean impact forces evaluated during the experiments .
Normally Dummy are utilized in crash test to have data from real incident among car.
We retain the original diction "Dummy" in order to better understand the use of the crash test method, but in fact in this research the "Judo Boy Dummy" is referred to as the reference "judo boy " which is defined as the average size of the children playing judo
In medical point of view, the result is similar to the "Reference Man" approach for Radio-Protection [34], it is a human being of statistically average size and physiology, used in research models of nutrition, pharmacology, population, radiologic protection and so on,.
Then we use "Judo Boy Dummy" for "Reference Judo Children" and utilize the accepted meaning in the Crash test methodology for similarity.
And following the **Crash Tests Methodology**, on this **Dummy** will be applied the average stress produced by each technique.
Thereby may be used formulas validated in Crash Tests Methodology, and we will get the consequent hazard results for children, associated with falls produced by judo throwing techniques.

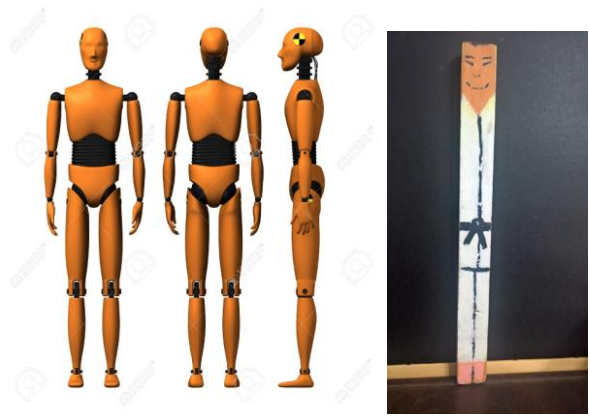

*Fig 19 20 from Crash test Dummy to Judo boy Dummy: "Gennaro Kano"*

Main Data for Judo Boy Dummy
Age = 11,8  y
h = 158  cm
W= 47   Kg
BSA =1,4495  m²
Main Contact Body Surface for each technique , with  average Uke body mass
MBSC1 =  0,0060   m²       m= 54,7 Kg
MBSC2 =  0,0093  m²        m = 53    Kg
MBSC3 =  0,0092  m²        m = 47,2  Kg
MBSC4 =   0,0115   m²       m = 49,4  Kg
MBSC5 =  0,0099  m²         m = 48,6  Kg

## 12) *Head Injury Criterion, Skull fracture Probability, Thoracic Trauma Index, Compression Criteria. Results for "Judo boy Dummy".*

In the crash Methodology there are defined and accepted some index that are the accepted connection between Impact Biomechanics and hazard or physiological injury, they are Head Injury Criterion, that depends directly from the head acceleration and impact time. This Criterion is directly connected to the probability a potential skull fracture.

$$HIC = max\left[\frac{1}{t_2 - t_2}\int_{t_1}^{t_2} a(t)dt\right]^{2,5} (t_2 - t_1) \quad (15)$$

$$p(fracture) = N\left(\frac{\ln(HIC - \mu)}{\sigma}\right) \quad (16)$$

The other important index is the Thoracic Trauma Index, that is based as important parameter on the Thorax acceleration multiplied by the actual body mass divided the average man body mass, with a connection to the age of subject.[35] This index it is not directly connected to the hazard of impact biomechanics but is an indicative index of the potential blunt traumas.

$$TTI = 1,4 AGE + 0,5(RIB_y + T12_y)(M/M_{std}) \quad (17)$$

More physiological connected is the AIS index ( Abbreviated Injury Scale ) that well is connected to the Compression criteria.
Defining compression (C) as the chest deformation divided by the thickness of the thorax.

$$AIS = -3,78 + 19,56 C \quad (18)$$

Evaluating with the mechanical and thermal data the Crash Test indexes for the "Judo Boy Dummy" we can single out, if judo is, generally speaking. a safe sport for children or not.

TTI = 26,4 ; 22,3 ; 26,5 ; 26,7 ; 21,7 <<80  No Traumatic Event.
AIS=-3,74; -3,76; -3,76; -3,76; -3,76.  <<0
Appling the theoretical compression 2mm in place of 0,5mm
AIS(0) ≤0 ( Extremely  Lightweight  Blunt Trauma).

## 13) *Hazard Training Sentinel : a Judo Kid training Digital Assistant.*

The two last subtle troubles about kids during still standing training of judo throwing techniques, with specific attention to Uke ( the receiver ) safety, are:
1. The wrong combination of boys during training of throwing techniques
   (i.e. to pair a very strong boy with a more delicate one)
2. And potential long-term traumas. (*blunt traumas potentially for Liver and Spleen* )

Trauma is one of the leading cause of morbidity and mortality in the pediatric population.
The abdomen, from medical statistics, is the third most commonly injured anatomic region in children, after the head and the extremities. [36][37]
Abdominal trauma can be associated with significant morbidity and may have a mortality as high as 8.5%. The abdomen is the most common site of initially unrecognized fatal injury in traumatized children.
Obviously the abdominal wall of a child has thinner musculature than that of an adult.
The ribs are more flexible, which makes them less likely to fracture; but, this increase in compliance makes them less effective at energy dissipation and, therefore, less effective at protecting the upper abdominal structures (eg, the spleen and the liver).[38]
Abdominal organs are comparatively larger in the child than in the adult; therefore, more surface area is exposed, making the organ more at risk for injury
From medical data more than 80% of traumatic abdominal injuries in children result from blunt mechanisms; most commonly, they are related to motor vehicle accidents. [39]
Abdominal injuries may also result from falls with blows to the abdominal wall.
The child's spleen has a thicker capsule than that of the adult, yet the spleen is among the most commonly injured solid organs in blunt abdominal trauma.
The problem rises not only by a direct blow from a fall produced by a stronger friend, but , in safety vision, also by the accumulation of micro-trauma into the two major target organs: Liver and Spleen. [40]
To help judo teachers in these *two important but under valuated* aspects of children safety. The software house ISS ( Italian Software Solutions ) is preparing a Digital Assistant ,on phone, PC, I-phone and Tablet based on this research, applied to all judo throws, so as to address these issues with a sound objectivity:
The name of the software is *Hazard Training Sentinel.*

*Fig21 HTS Phone App*

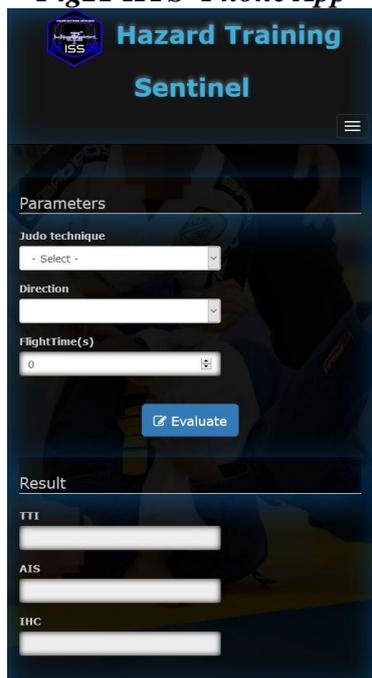

This software is able not only to evaluate the stress produced by a single throw, for each boy, on the left and on the right, but also to evaluate the micro-trauma in time and to alert the teacher when the kids /child needs to stop for sometime falling down, to give recover time to children body playing judo without falls. Probably these finding will change thye basic training for kids, with the introduction of without fall recovery time
.

## *14) Conclusions.*

The soul of this work is to build a sound methodology grounded on scientific criteria to evaluate the hazard in judo throws for children.
Because it is not possible to connect impact mechanics to the resulting traumas, in a clear way; it is introduced the "Judo Boy Dummy" to utilize verified methods (crash tests) to do so.
The experimental measurements of flight times and body contact area for each falls, as input for the hazard evaluation, clearly show that for falls produced by judo throws among boys:
 **Judo is a Safe Sport For Children.**
  *The hazard training sentinel will be a friendly and useful software support for teachers, to teach throws in safe way, during the first years of kids training*

## *15) References.*

[1] David Geier *Sport Medicine Simplified* Dr. David Geier Enterprises, LLC BN ID: 2940148242253
[2] S.Salanne, & coworkers *Traumatismes secondaires à la pratique du judo chez l'enfant* Atrchives de Pediatrie 2009 XXX, 1-8.
[3] Barsottini & coworkers *Relationship between techniques and injuries among judo practitioners* Rev. Bras. Med. Deporte vol 12 N° 1 , 2006
[4] Pocecco E, & coworkers . *Injuries in judo: a systematic literature review including suggestions for prevention* Br J Sports Med 2013;47:1139–1143. doi:10.1136/bjsports-2013-092886
[5] H. E. Roland & B. Moriarty *System Safety Engineering and Management, Second Edition*. Copyright © 1990 John Wiley & Sons, Inc. ISBN 0-471-61816-0
[6] Hollnagel E., Wears R.L. and Braithwaite J. *From Safety-I to Safety-II: A White Paper. The Resilient Health Care Net*: Published simultaneously by the University of Southern Denmark, University of Florida, USA, and Macquarie University, Australia. 2015
[7] Kwon, Kim, Cho *"A Kinematics Analysis of Uchi-mata (inner thigh reaping throw) by Kumi-kata Types and Two Different Opponents Height in Judo (2)"* Korean journal of sports biomechanics 2005
[8] Kim, Yon *"A kinematic analysis of the attacking-arm-kuzushi motion as to pattern of morote-seoinage in judo"* Korean Journal of Sport Biomechanics Vol13 2003 N°1
[9] Ishii and Ae : *Biomechanical factor of effective Seoi Nage in Judo* Doctoral program in Physical Education Fitness and Sport Science Tsukuba Japan 2014
[10] Imamura, R.T., Hreljac A., Escamilla R.F., Edwards W.B. *A Three-Dimensional Analysis Of The Center Of Mass For Three Different Judo Throwing Techniques*. Journal of Sports Science and Medicine CSSI, 122-131, 2006.
[11] Da Costa J.M., E. *Analise Biomecanica da Tecnica de Judo – Sasae-Tsuri-Komi-Ashi.* Dissertacao apresentada a prova de mestrado no ramo das ciencias do desporto, espaciallidade de treino de alto rendimento, nos termos do capitulo II do Decreto-Lei n° 216/92 de Outubro, Porto, Universidade do Porto, Faculdade De Ciencias do Desporto e de Educacao Fisica, 2003.

[12] Suarez, G.R. *Análisis De Factores Biomecánicos Y Comportamentales Relacionados Con La Efectividad Del Uchi Mata, Ejecutado Por Judokas De Alto Rendimiento.* Universidad De Granada, Tesis Doctoral, Edución Física Y Deportiva, Ciencias De La Actividad Física Y El Deporte.
[13] Ibrahim Fawzi Mustafa, *Force impulse of body parts as function for prediction of total impulse and performance point of Ippon Seoi Nage skill in judo* World Journal of Sport science N° 3 , 2010
[14] Blais, Trilles : *Analyse méchanique comparative d'une meme projection de Judo: Seoi Nage, realisée par cinq experts de la Fédération Francaise de Judo*. Science et Motricité N°51, 2004
[15] Blais, Trilles, Lacoture : *Détermination des forces de traction lors de l'exécution de Morote Seoï Nage réalisée par 2 experts avec l'ergomètre de Mayeur et un partenaire* Journal of Sport Science 2007
[16] Stronge *Impact mechanics* Cambridge UniversityPress 2004 ISBN 0-521-60289-0
[17] Hitosugi & coworkers *Biomechanical analysis of acute subdural hematoma resulting from judo* Biomedical Research (Tokyo) **35** (5) 339-344, 2014
[18] Murayama & coworkers *Rotational Acceleration during head impact resulting from different judo throwing techniques*. Neuro Med Chir ( Tokyo) 54, 2014
[19] Ionescu M. *Chemistry and Technology of Polyols for Polyurethanes, 2nd Edition Volume I&II* Smithers Rapra Publisher 2016 ISBN: 978-1-91024-213-1
[20] Mane & coworkers *Mechanical Property Evaluation of Polyurethane Foam under Quasi-static and Dynamic Strain Rates- An Experimental Study* Procedia Engineering 173 ( 2017 ) 726 – 731
[21] Bradley & Sullivan *Thermal Expansion of Polyurethane Foam* 43rd Annual Technical Meeting of the Society of Engineering Science The Pennsylvania State University August 2006
[22] Sychev *Complex Thermodynamic Systems (Studies in Soviet Science)* Consultant bureau New York 1973.
[23] Latif *Heat Conduction* Springer 2009 ISBN 978-3-642-01266-2
[24] Quesada *Application of infrared thermography in Sport Science* Springer 2017 ISBN 978-3-319-47409-0
[25] Sacripanti & coworkers *Infrared Thermography- Calorimetric Quantitation of Energy Expenditure in Biomechanically Different Types of Jūdō Throwing Techniques. A Pilot Study* Annals of Sport Medicine and Research 2015
[26] Kai-Uwe Schmitt &coworkers *Trauma Biomechanics-accidental injuries in traffic and sports* Springer 2007 ISBN 978-3-540-73872-5
[27] Kai-Uwe Schmitt &coworkers *Trauma Biomechanics-introduction to injury biomechanics* Springer 2014 ISBN 978-3-642-53919-0
[28] Fung & coworkers *Accidental Injury Biomechanics and Prevention* Springer 2002 ISBN 978-1-4419-3168-9
[29] Yoganandan & coworkers *Accidental Injury Biomechanics and prevention* Springer 2015 ISBN 978-1-4939-1731-0
[30] Franck & Franck *Forensic biomechanics and Human Injury* CRC Press 2016 ISBN 13: 978-1-4822-5888-2
[31] Pilkey & coworkers *Injury Biomechanics and Control* Wiley & Sons 2010 ISBN: 978-0-470-10015-8.
[32] Visvikis & coworkers *Child safety: Q-Series crash test dummy family regulatory application assessment* EU Final Report January 2015
[33] Kuppa *Injury criteria for side impact dummies* National Transportation Biomechanics Research Center, National Highway Traffic Safety Administration May 2004.

[34] ICRP *Report on the Task Group on Reference Man* ICRP Publication 23 1975
[35] Thunissen & coworkers *Scaling of adult to child responses applied to the thorax* TNO Crash Safety research Center the Nederland 1994
[36] Bilo & coworkers *Forensic aspect of pediatric fractures* Springer 2010 ISBN: 978-3-540-78715-0
[37] Yong-Ping Zheng & Yan-Ping Huang *Measurement of soft Tissue Elasticity in Vivo* CRC Press 2016 ISBN 13: 978-1-4665-7629-2
[38] Chavez & Mendoza *Soft Tissue Composition injury mechanism and repair* Nova Science Publisher 2012 ISBN: 978-162257-371-4
[39] Franck and Franck *Forensic biomechanics and human injury* CRC Press 2016 ISBN 13: 978-1-4822-5888-2
[40] Avril *Material Parameter Identification and Inverse Problems in Soft Tissue Biomechanics* Springer 2017 ISBN 978-3-319-45070-4